\documentclass{article}
\usepackage{frascatiphys}
\usepackage{epsfig}
\newcommand{\subrm}[1]{\mbox{\tiny \rm #1}}

\newcommand{\Br}{{\rm Br}}
\newcommand{\epe}{\epsilon'/\epsilon\,}
\newcommand{\Reepe}{\rm{Re(}\epsilon'/\epsilon \rm{)}}

\newcommand{\kl}{K_{\subrm{L}}}
\newcommand{\ks}{K_{\subrm{S}}}

\newcommand{\ksl}{K_{\subrm{S,L}}}

\newcommand{\klpiee}{\kl \to \pi^0 e^+ e^-}

\newcommand{\klpipipi}{\kl \to 3 \pi^0}

\newcommand{\klpizpiz}{\kl \to 2 \pi^0}
\newcommand{\kspizpiz}{\ks \to 2 \pi^0}

\newcommand{\ksgg}{\ks \to \gamma \gamma}
\newcommand{\klgg}{\kl \to \gamma \gamma}

\newcommand{\kslgg}{\ksl \to \gamma \gamma}
\newcommand{\kspigg}{\ks \to \pi^0 \gamma \gamma}
\newcommand{\klpigg}{\kl \to \pi^0 \gamma \gamma}

\begin{document}
\title{ 
TESTS OF CHIRAL PERTURBATION THEORY IN 
RARE KAON DECAYS
}
\author{
Rainer Wanke        \\
{\em Institut f\"ur Physik, Universit\"at Mainz, D-55099 Mainz, Germany} \\
}
\maketitle
\baselineskip=11.6pt
\begin{abstract}
The neutral Kaon decays $\ksgg$ and $\klpigg$ are very sensitive to higher order
loop effects of Chiral Perturbation Theory (ChPT). 
New measurements of the NA48 experiment show that ChPT contributions of ${\cal O}(p^6)$
cannot be neglected in these modes.
In addition a new measurement of the related decay $\klgg$ and an upper limit 
of the rate of $\kspigg$ are presented.
\end{abstract}
\baselineskip=14pt

\begin{figure}[t]
\begin{center}
\epsfig{file=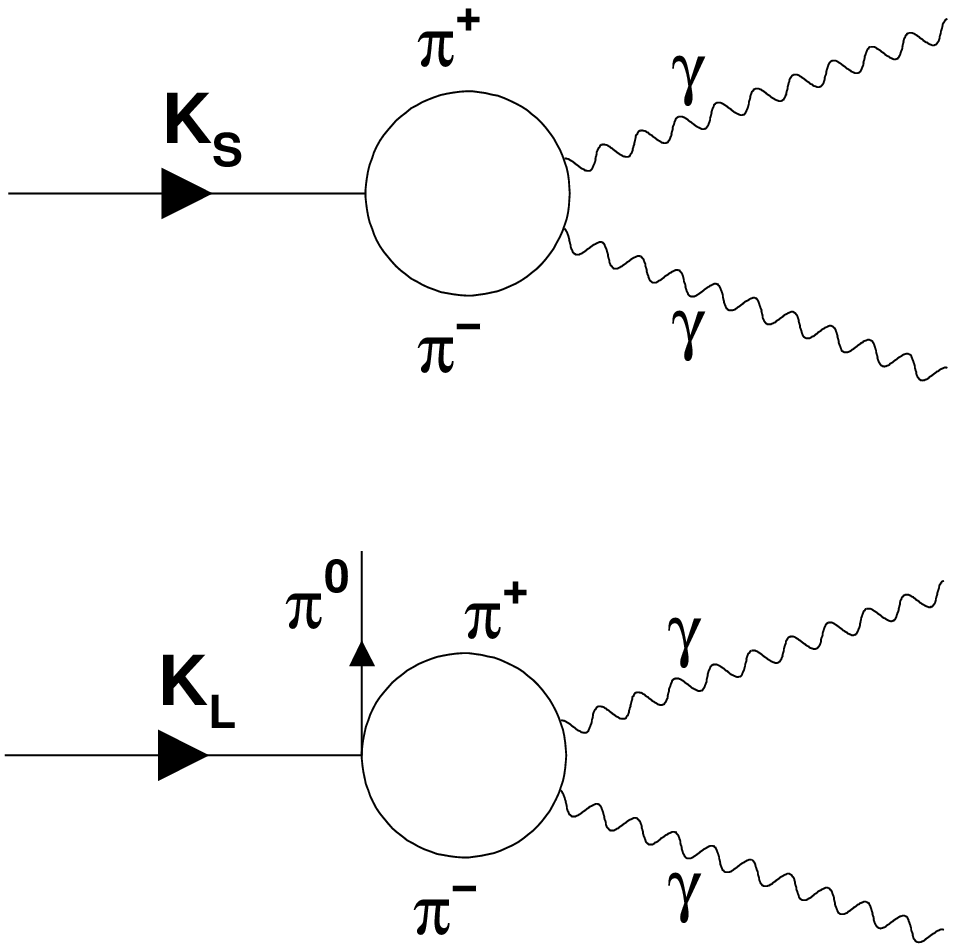,width=0.32\textwidth}
\hfill
\epsfig{file=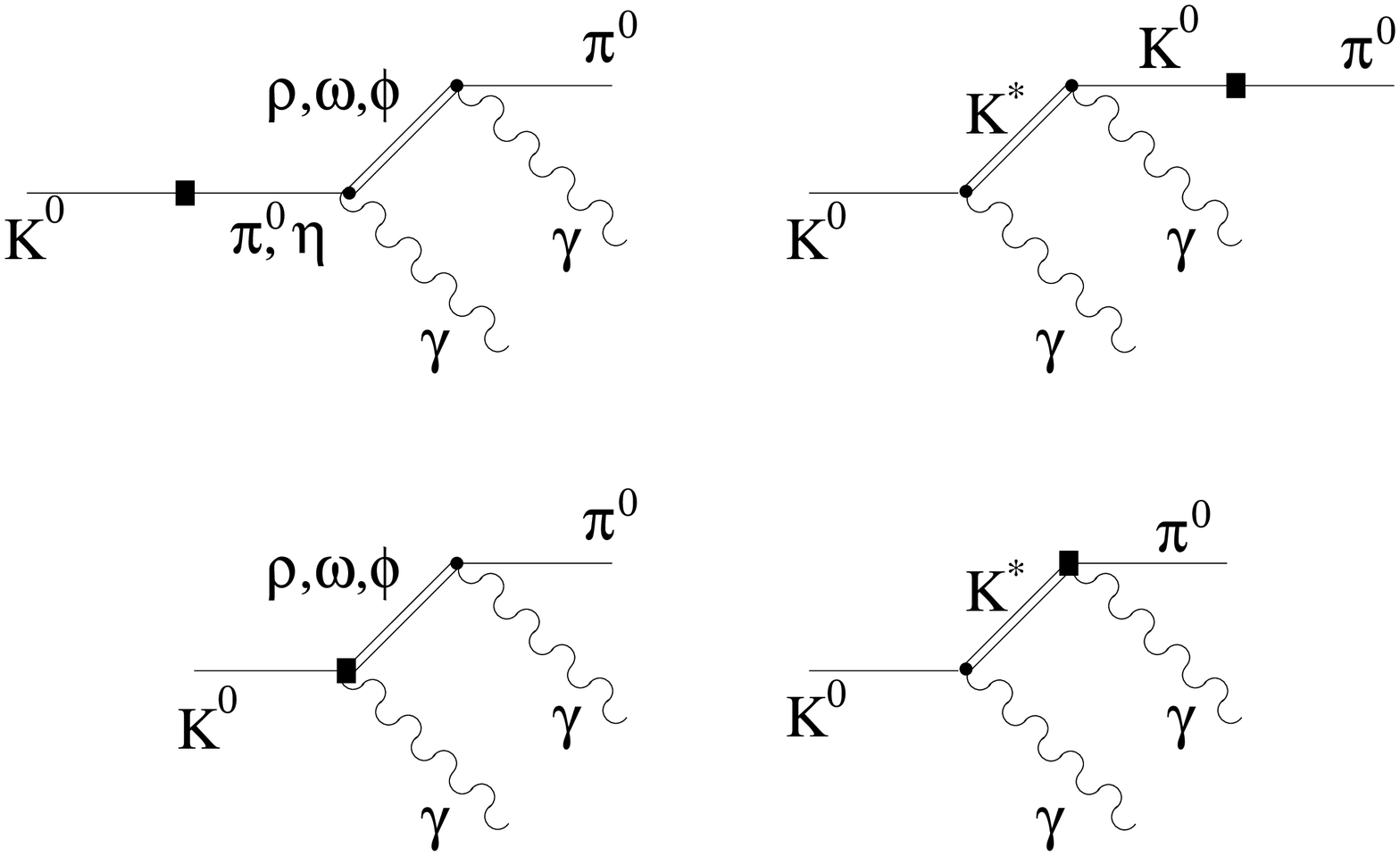,width=0.58\textwidth}
\end{center}
\vspace{-4mm}
\caption{\it Left:Examples of ${\cal O}(p^4)$ loop diagrams for $\ksgg$ and $\klpigg$.
Right: Vector meson exchange diagrams for $\klpigg$.
\label{fig:klpigg_feynman} }
\end{figure}

\section{Introduction}

Chiral Perturbation Theory (ChPT) has been proven as an extremely successful
effective theory for low energy hadron dynamics.
Nevertheless, the knowledge of higher order effects is rather scarse at present.
The decays $\ksgg$ and $\klpigg$ are well suited for investigation of
higher order contributions. In both cases the lowest order ${\cal O}(p^2)$
vanishes\cite{bib:gaillardlee} and the next order ${\cal O}(p^4)$ can precisely been calculated,
since no counter-terms exist\cite{bib:ksgg_theo, bib:klpigg_theo} (Fig.~\ref{fig:klpigg_feynman} (left)).

The ${\cal O}(p^4)$ prediction of the $\ksgg$ branching ratio is
$\Br(\ksgg) = 2.1 \times 10^{-6}$ to an accuracy of about $5\%$\cite{bib:ksgg_theo}.
Calculations of the next order ${\cal O}(p^6)$ do not exist up to now.
The measured value of $\Br(\ksgg) = (2.6 \pm 0.5) \times 10^{-6}$\cite{bib:ksgg_na48}
is in agreement with this prediction, however, the experimental errors were still too large to allow an accurate comparison.


For $\klpigg$ it is known that ${\cal O}(p^4)$ alone underestimates
the observed branching fraction by about a factor of three\cite{bib:klpigg_theo, bib:klpigg_prev, bib:klpigg_ktev}. 
At ${\cal O}(p^6)$ the rate can be reproduced by adding a
vector meson exchange contribution\cite{bib:vmd} (Fig.~\ref{fig:klpigg_feynman} (right)) via the coupling constant $a_V$\cite{bib:avtheo}.
However, the parameter $a_V$ has to be measured experimentally.
Additional interest in measuring $\klpigg$ arises, since it can constrain 
the CP conserving amplitude of the direct CP violating decay $\klpiee$.
The VMD mechanism could enhance the size of this amplitude, depending on the value
of $a_V$\cite{bib:klpiee_theo}.

\section{Experimental Method}

With KTeV and NA48 mainly two experiments have investigated neutral kaon decays into neutral final
states in the recent years. Both experiments were built to perform precise measurements
on the parameter $\epe$ of direct CP violation.
Both are fixed target experiments with the neutral kaon beams being produced by
high-energetic proton beams.

\subsection{The NA48 Experiment}

The NA48 experiment, located at the SPS accelerator at CERN, has been taking data
for the $\Reepe$ measurement in the years $1997 - 1999$ and 2001.
In addition to the regular $\epe$ data taking, several runs with
a high intensity $\ks$ beam have been taken, where the $\ks$ intensity has been
increased by more than a factor of 200 with respect to the $\epe$ runs.
In the year 2000, the spectrometer has not been available. In this year half
of the data taking took place under $\epe$ conditions while the second half
of the run period was performed with a high intensity $\ks$ beam.

\subsection{The KTeV Experiment}

The KTeV experiment is located at the TeVatron at Fermilab.
It has been taking data in the years 1996, 1997, and 1999 with runs
dedicated to determine $\Reepe$ and runs for measuring rare $\kl$ and
hyperon decays. The data from the last year of data taking are 
mostly still being analyzed.

\subsection{Reconstruction of Neutral Decays}


For detecting the photons in the final states of the decays discussed here,
NA48 and KTeV use a quasi-homogeneous liquid krypton and, respectively, a CsI crystal
electromagnetic calorimeter.
By assuming a kaon decay, the $z$ position (along the beam pipe)
of the decay vertex can be calculated to
\[
z_{\subrm{vertex}} \; = \; z_{\subrm{calorimeter}} \: - \: \frac{1}{m_K} \, \sqrt{\, \sum_{i>j} E_i E_j d_{ij}^2} \, ,
\]
with the shower energies $E_{i,j}$, distances $d_{ij}$, and the nominal kaon mass $m_K$.
If photons are lost, the missing energy shifts the vertex position down-stream towards the calorimeter.
If the decay contains one or more intermediate $\pi^0$, background can be suppressed by requiring
the $\pi^0$ decay vertex to be consistent with the kaon decay vertex.

\section{$\kslgg$}

For a decay rate measurement of $\ksgg$
in a fixed-target experiment an irreducible background
of $\klgg$ events has to be subtracted.
Therefore, a precise determination of $\klgg$
is necessary while the current world average on $\Br(\klgg)$ has a relative
error of about $3\%$\cite{bib:pdg}.
For this reason the NA48 experiment has used a different method for subtracting
$\klgg$ events: In using the $\kl$ target run in 2000 with a very similar experimental set-up
the relative rate $\Gamma(\klgg)/\Gamma(\klpipipi)$ is measured. By also measuring the
$\klpipipi$ rate in the high intensity $\ks$ run 2000, the number of background $\klgg$ events
can then be accurately subtracted.

\subsection{$\klgg$}

\begin{figure}[t]
\begin{center}
\epsfig{file=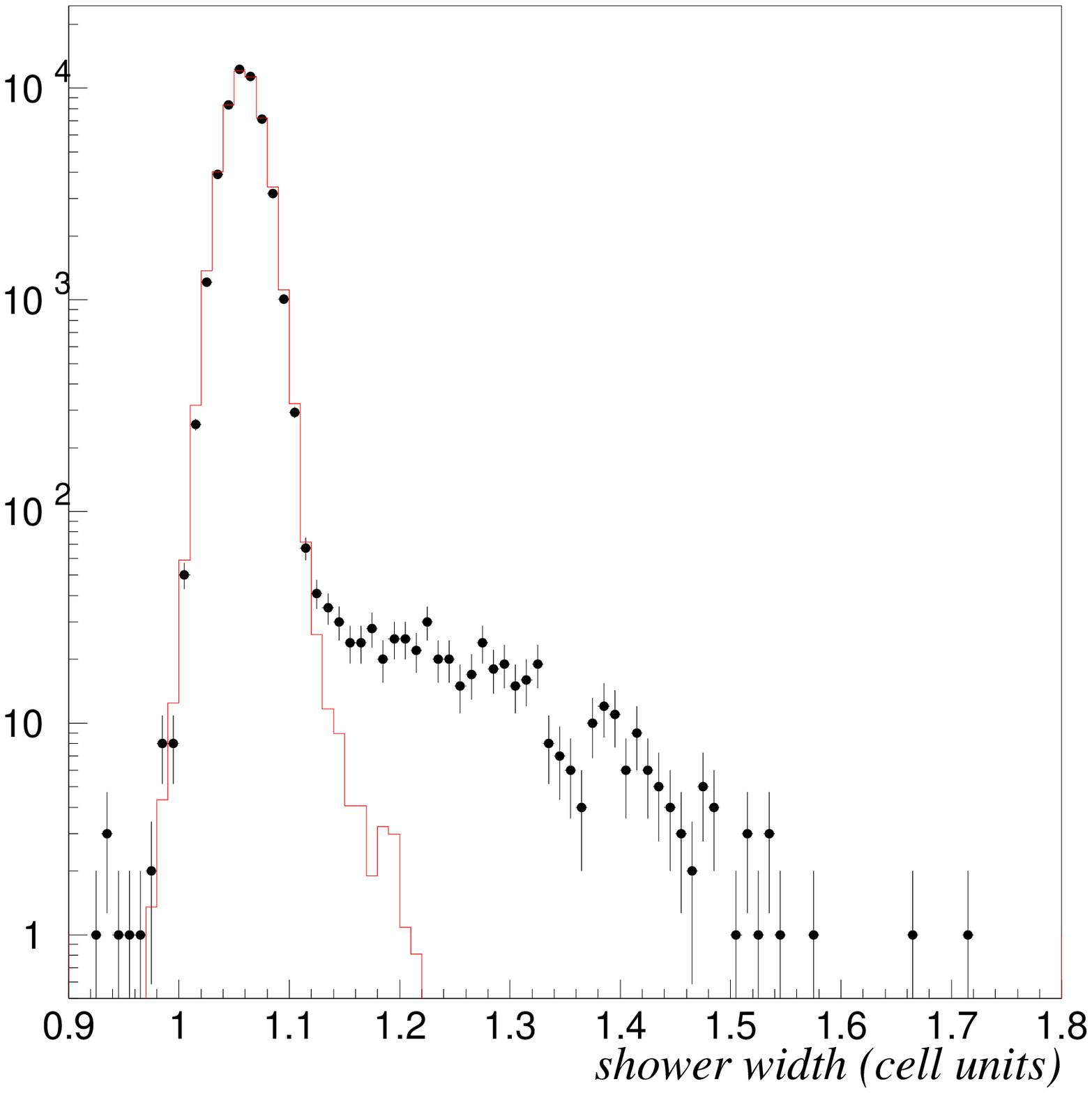,width=0.4\textwidth}
\hspace{0.05\textwidth}
\epsfig{file=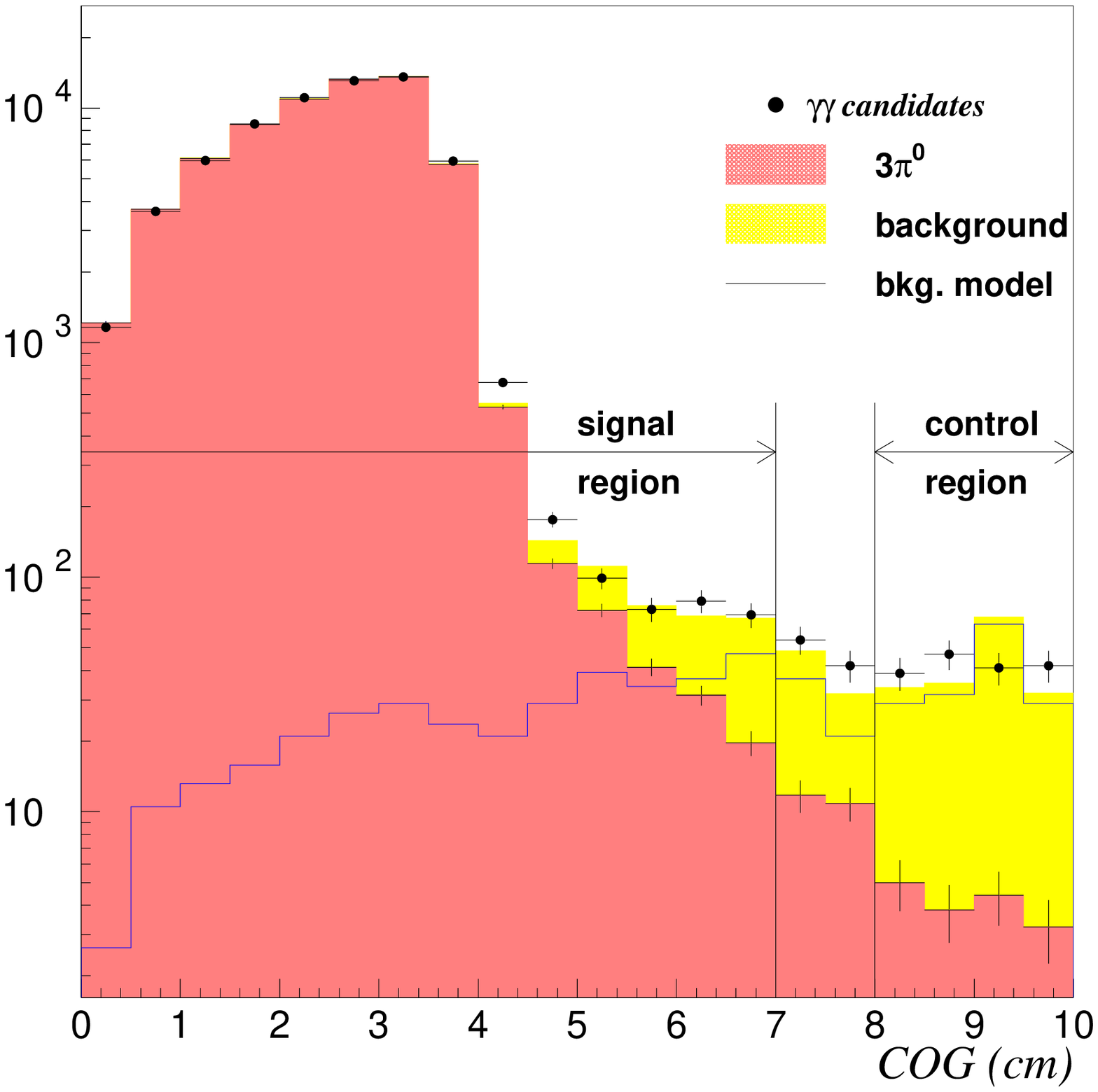,width=0.4\textwidth}
\end{center}
\vspace{-6mm}
\caption{\it Shower width (left) and center-of-gravity distribution (right) of the NA48 $\klgg$ candidates
compared with $\klpipipi$ events.
\label{fig:klgg_bkg} }
\end{figure}

The NA48 $\kl$ target run of the year 2000 has provided very clean conditions
to measure $\klgg$ decays. Backgrounds from $\klpizpiz$ or other $\kl$ decays
are completely negligible, in particular $\kl \to e^+ e^- \gamma$ Dalitz decays are
swept out by the spectrometer magnet.
The only remaining background source is hadronic activity,
e.g.\ from $\Lambda$ decay products, which in rare cases 
might enter the decay volume via the $\kl$ beam line.
To estimate this background the sidebands in the shower width and
center-of-gravity distributions are evaluated (Fig.~\ref{fig:klgg_bkg}).
By using both methods, the hadronic background is determined to
$(0.6 \pm 0.3)\%$, where the error reflects the difference
of the two estimations.

The normalization channel $\klpipipi$, which needs three good $\pi^0 \to \gamma \gamma$ combinations
to be selected, is virtually background-free.
Both, signal and normalization channel have trigger efficiencies larger than $99\%$.
For the analyis only $25\%$ of the $\kl$ target run data were used which already provides sufficient statistics.
Evaluating the event numbers in the vertex region $-1$~m~$< z < 5$~m (measured from the $\ks$ collimator)
and taking the acceptance ratio from Monte Carlo simulation, NA48 finds
\[
\frac{\Gamma(\klgg)}{\Gamma(\klpipipi)} \; = \; ( 2.81 \, \pm \, 0.01_{\subrm{stat}} \, \pm \, 0.02_{\subrm{syst}} ) \times 10^{-3}
\]
The systematic error is dominated by uncertainties in the $\gamma \gamma/ 3 \pi^0$ acceptance ratio.
This result improves the current world average by about a factor of four.

\subsection{$\ksgg$}

In addition to the irreducible $\klgg$ decays more background sources have to be taken
into account when selecting $\ksgg$ candidates:

$\kspizpiz$ with lost and/or overlapping photons may fake a good $\gamma \gamma$ event.
Since in these cases energy is lost, the neutral vertex is shifted down-stream.
Background from $\kspizpiz$ can therefore be efficiently rejected by restricting the
allowed vertex range to be within $-1$ and 5~m behind the final collimator. 
Doing so, the remaining background from $\kspizpiz$ is estimated to $(0.8 \pm 0.2)\%$,
where the uncertainty is arising from the shower overlap probabilities being different in
the simulation as in the data.

\begin{figure}[t]
\begin{center}
\epsfig{file=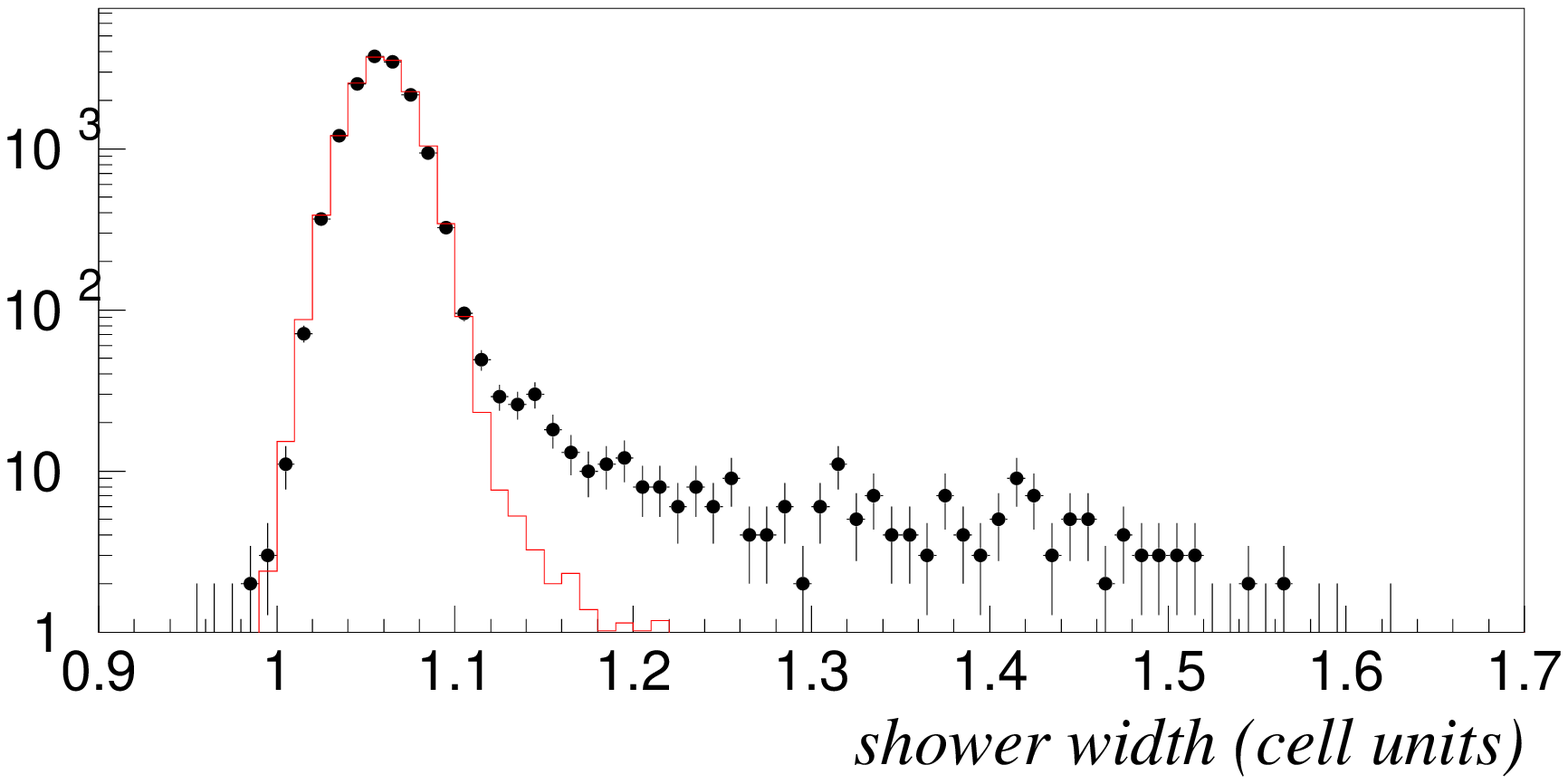,width=0.6\textwidth,height=0.255\textheight}
\epsfig{file=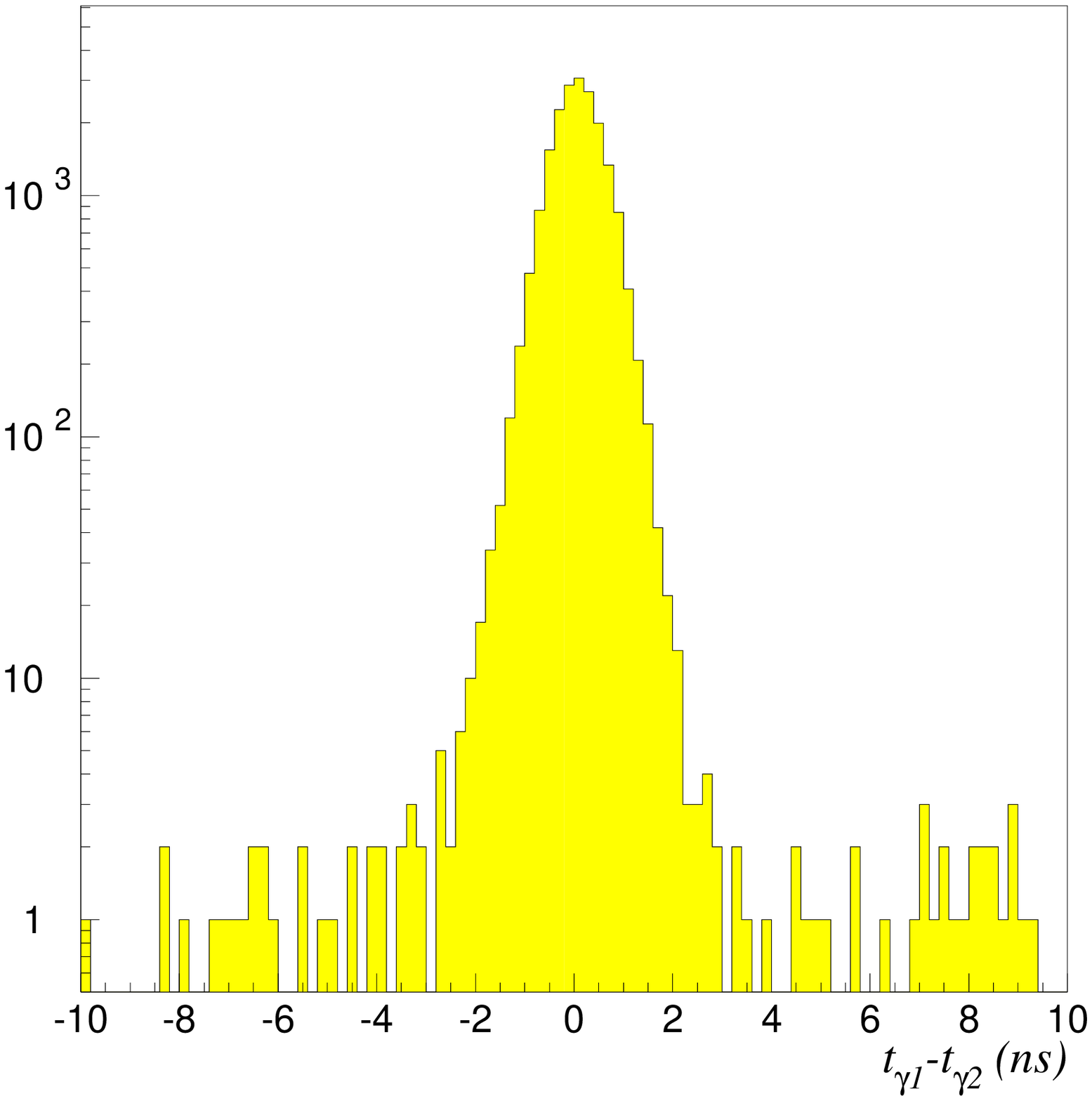,width=0.39\textwidth,height=0.23\textheight}
\end{center}
\vspace{-6mm}
\caption{\it Distributions of the normalized shower width (left) and the cluster time difference (right)
of the NA48 $\ksgg$ candidates compared with $\klpizpiz$ events.
\label{fig:ksgg_bkg} }
\end{figure}

Further background sources are hadronic events (originating e.g.\ from scattering at the 
collimator) or from accidentally overlapping events.
Both are determined by a sideband subtraction in the center-of-gravity distribution of
the LKr calorimeter (Fig.~\ref{fig:ksgg_bkg}).
From this, the total background from hadronic and accidental activity is estimated
to $(0.8 \pm 0.3)\%$.

Finally, Dalitz decays $\pi^0 \to e^+ e^- \gamma$ and $K^0 \to e^+ e^- \gamma$
have to be taken into account. In the high intensity $\ks$ running in 2000, the NA48 experiment 
had no magnetic field in the detector. Therefore, many Dalitz $e^+e^-$ pairs did not separate
and were overlapping in the calorimeter. From Monte Carlo Simulation, the probability of
a Dalitz decay misidentified as $\gamma \gamma$ pair was estimated to about $30\%$.
With Dalitz decay probabilities assumed to be equal for $K^0 \to \gamma \gamma$ and $\pi^0 \to \gamma \gamma$,
the effect is twice as big for the $\kspizpiz$ normalization than in the $K^0 \to \gamma \gamma$
events and results in a relative correction to the $\ksgg$ branching ratio
of $(+1.5 \pm 0.3)\%$.

\begin{figure}[t]
\begin{center}
\epsfig{file=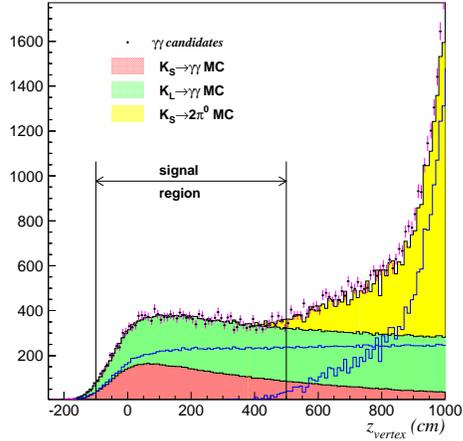,width=0.55\textwidth}
\end{center}
\vspace{-6mm}
\caption{\it $z$ vertex distribution of $\gamma \gamma$ candidates.
\label{fig:ksgg_zvertex} }
\end{figure}

The decay vertex distribution of the NA48 $\gamma \gamma$ candidates is shown in
Fig.~\ref{fig:ksgg_zvertex} together with the estimated contributions of
$\ksgg$, $\klgg$ and $\kspizpiz$ background. The $\kspizpiz$ background contribution
has been normalized to the absolute $\ks$ flux.
In the fiducial region between $-1$ and 5~m about 20000 $\gamma \gamma$ candidates were found.
Subtracting all background contributions and normalizing to fully reconstructed $\kspizpiz$
events, the branching fraction was determined to
\[
\Br(\ksgg) \; = \; ( 2.78 \, \pm \, 0.06_{\subrm{stat}} \, \pm \, 0.04_{\subrm{syst}} ) \times 10^{-6}.
\]
The systematic uncertainty is dominated by the branching fraction of the $\kspizpiz$ normalization
($\pm 0.9\%$), the estimation of the hadronic and accidental background ($\pm 0.7\%$)
and the Monte Carlo statistics ($\pm 0.6\%$).

This new result significantly improves the previous measurements and is in clear discrepancy with
the ${\cal O}(p^4)$ based theoretical predictions.

\section{$\klpigg$}

With four photons in the final state
the decay $\klpigg$ has almost the same signature as the
CP-violating decay $\klpizpiz$, which is one of the signal decays
for the $\Reepe$ measurement of the KTeV and NA48 experiments.
Therefore the decay $\klpigg$ can be taken in parallel with 
regular $\epe$ data taking.
Moreover, both signal and normalization ($\klpizpiz$)
have practically identical trigger efficiencies.
However, the analysis of $\klpigg$ events has to fight strong backgrounds,
in particular $\klpipipi$ events with lost and/or overlapping photons,
badly reconstructed $\klpizpiz$ events, and accidentally overlapping events.

\begin{figure}[t]
\begin{center}
\epsfig{file=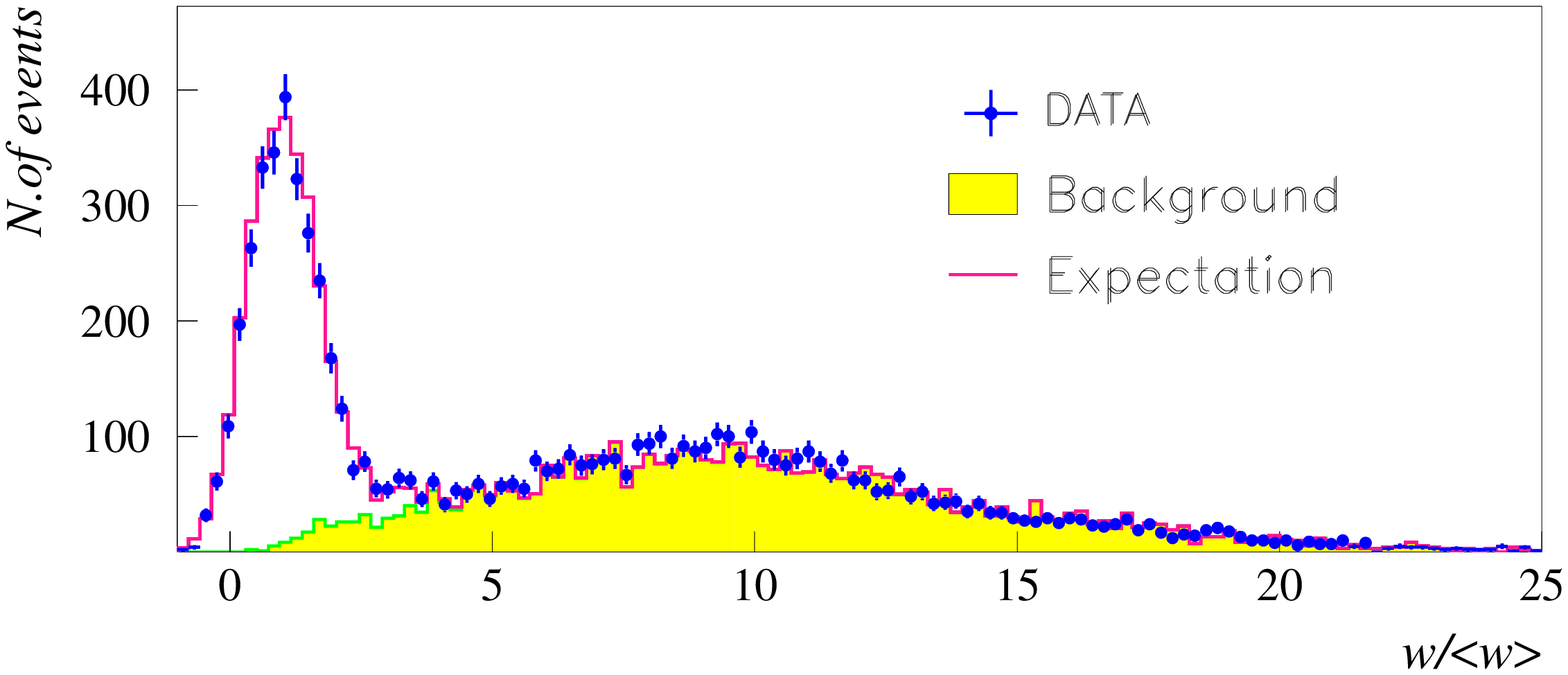,width=0.49\textwidth,height=0.2\textheight}
\epsfig{file=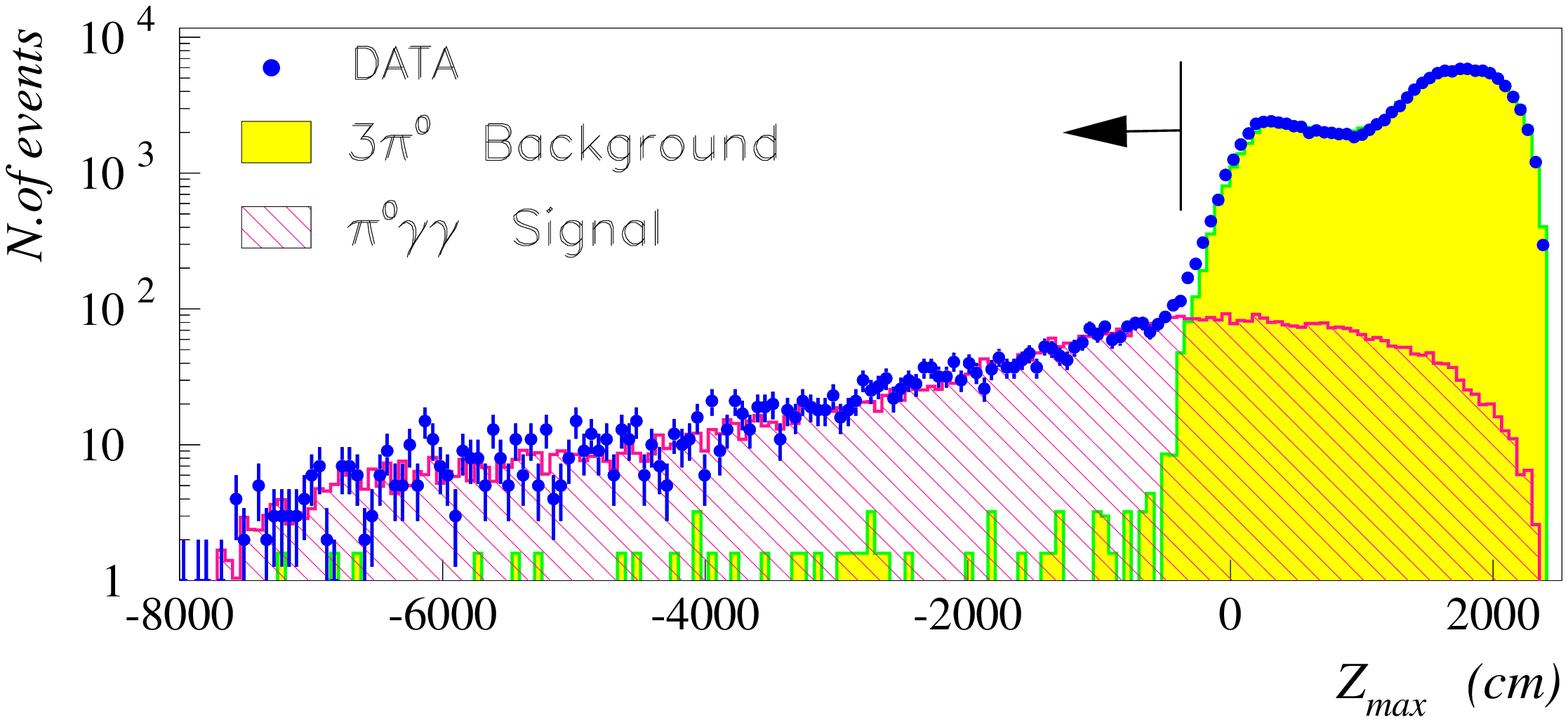,width=0.49\textwidth,height=0.2\textheight}
\end{center}
\vspace{-6mm}
\caption{\it Distributions of shower width (left) and $z_{\subrm{max}}$ (right)
of the selected $\klpigg$ candidates of the NA48 experiment.
Shown are the data (crosses), the $\klpipipi$ background expectation from Monte Carlo
simulation (shaded), and the total expectation (histogram).
\label{fig:klpigg_bkg} }
\end{figure}

The NA48 analysis\cite{bib:klpigg_na48}, which is described more detailed in the following, 
is based on the NA48 data of the years 1998 and 1999.
The suppression of background from $\klpipipi$ with overlapping or lost photons
is done by cutting on the shower width in the calorimeter.
As shown in Fig.~\ref{fig:klpigg_bkg} (left), 
overlapping photons produce on average a much wider shower and can clearly be
separated from the signal.
To suppress the $\klpipipi$ background even further,
a variable $z_{\subrm{max}}$ is constructed as estimate of the real kaon vertex
under the assumption of a misidentified $\klpipipi$ event.
For real background events this variable should be in the physical region down-stream
of the collimator (Fig.~\ref{fig:klpigg_bkg} (right)). By rejecting events with $z_{\subrm{max}} > -5$~m,
the $\klpipipi$ background is reduced to $(2.7 \pm 0.4)\%$.
However, the signal acceptance is also reduced by $54\%$.

A second source of background are misidentified $\klpizpiz$ events. They are
rejected by requiring the invariant mass $m(\gamma_3 \gamma_4)$ of the
photons not coming from the $\pi^0$ to be outside a window of $\pm 25$~MeV/$c^2$
around the nominal $\pi^0$ mass.
By using $\kspizpiz$ events from both $\epe$ and high intensity $\ks$ runs, the
remaining background from $\klpizpiz$ is estimated to $(0.2 \pm 0.1)\%$.

Finally, good $\pi^0 \gamma \gamma$ candidates may be mimicked by
accidentally overlapping events. This background is estimated by
extrapolating events with high transverse momentum into the signal region
to $(0.3 \pm 0.2)\%$.


\begin{figure}[th]
\begin{center}
\epsfig{file=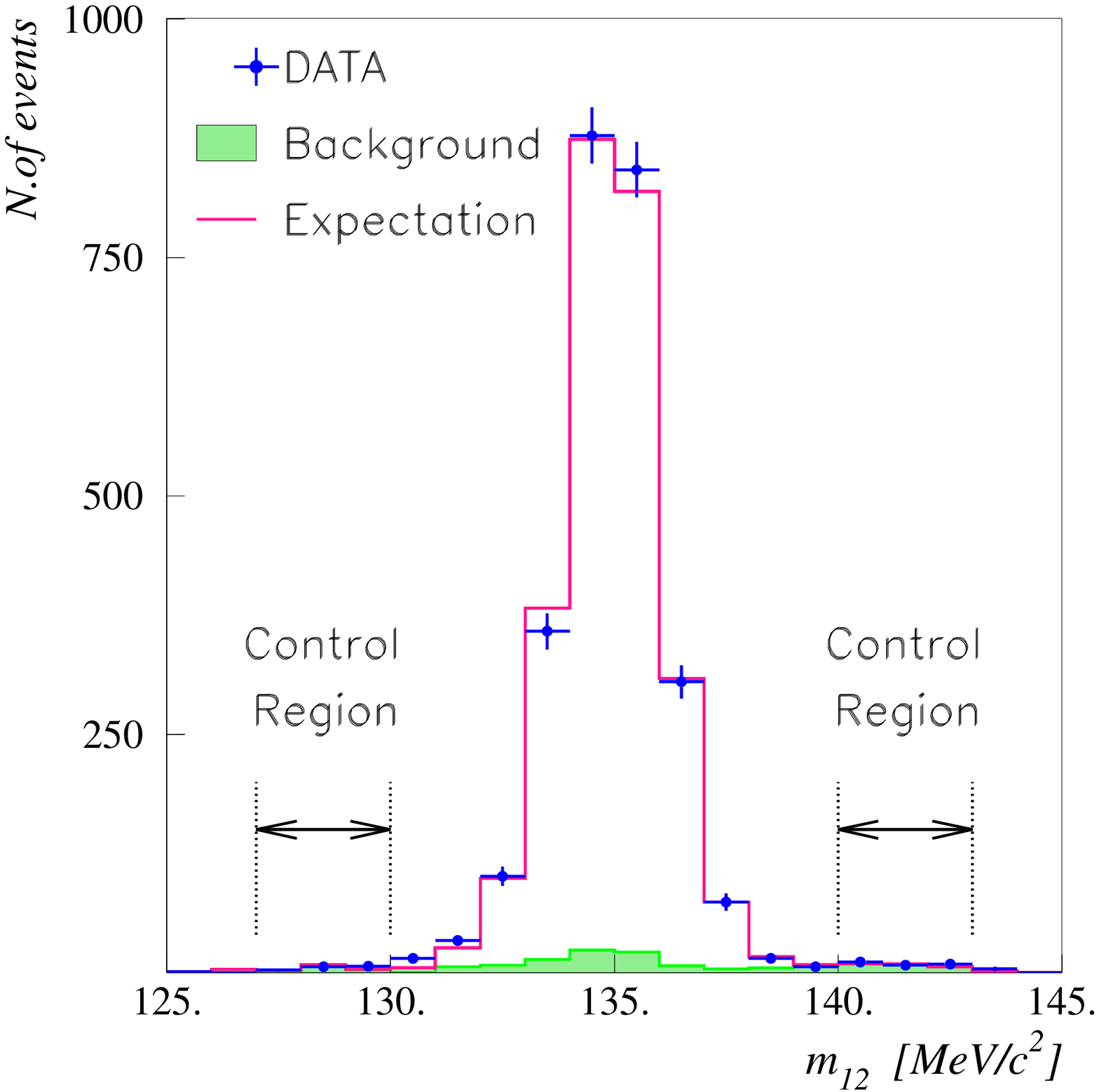,width=0.32\textwidth}
\epsfig{file=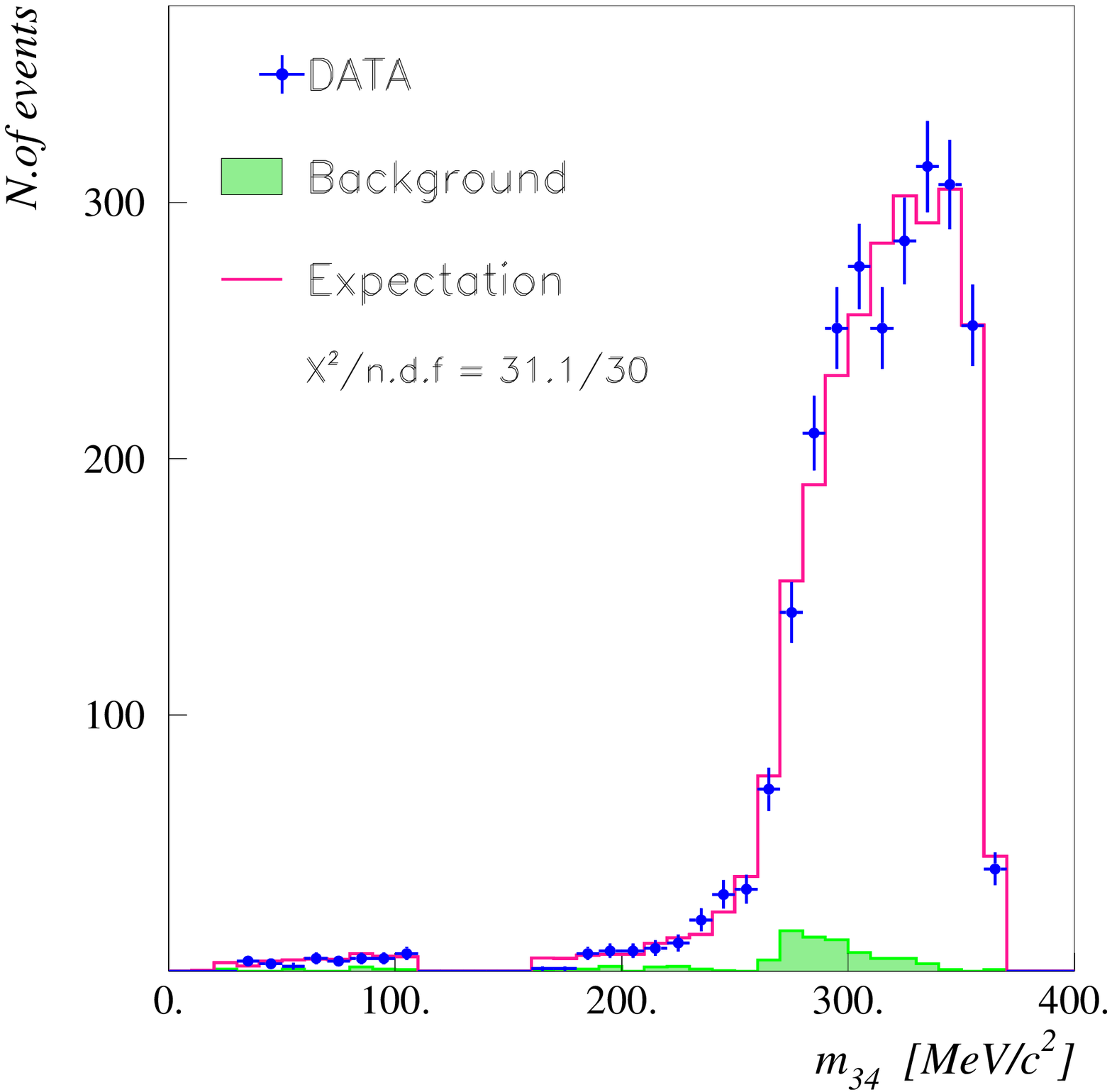,width=0.32\textwidth}
\epsfig{file=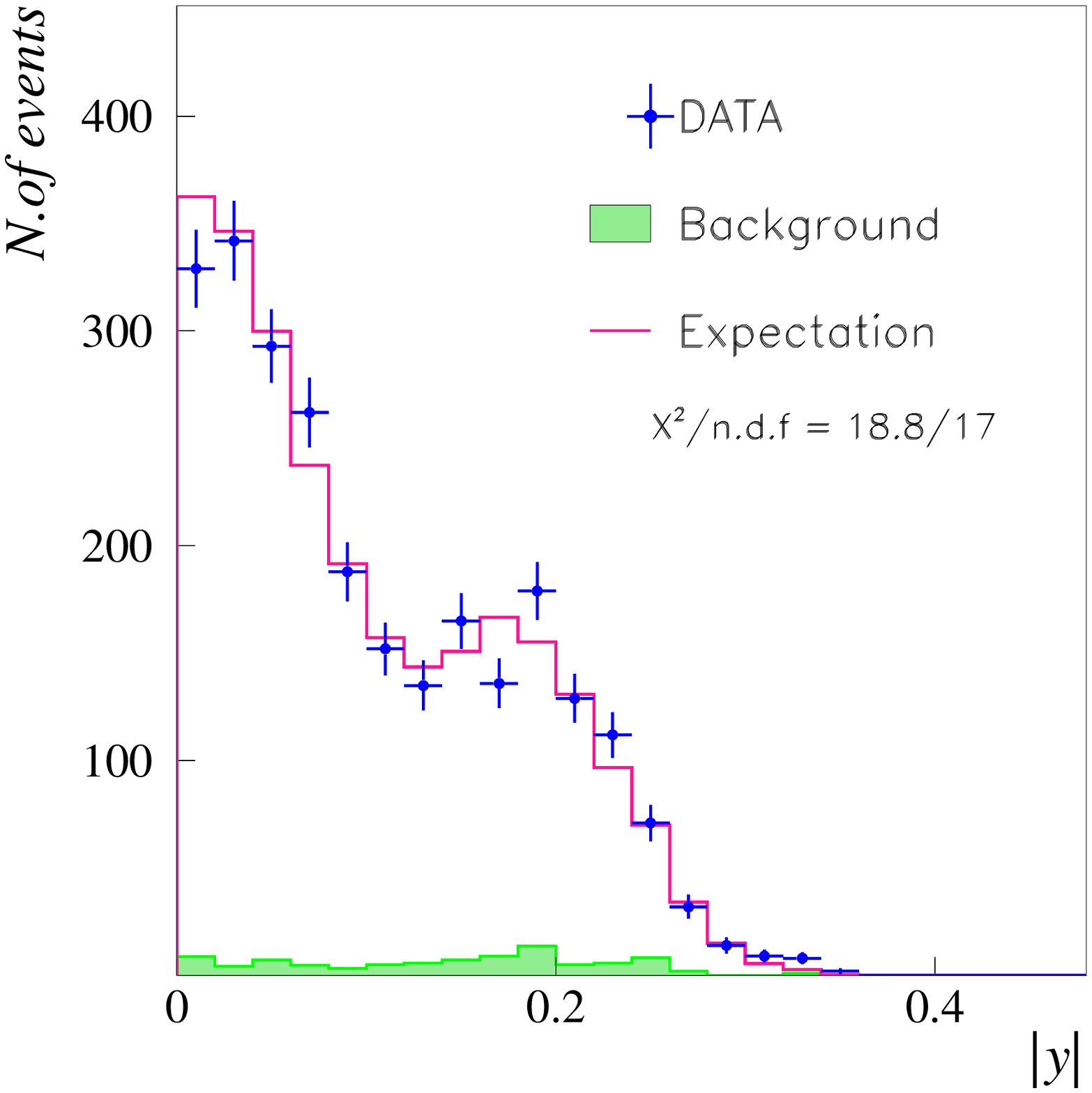,width=0.33\textwidth}
\vspace{-6mm}
\end{center}
\caption{\it Distributions of $m(\gamma_1 \gamma_2)$ (left), $m(\gamma_3 \gamma_4)$ (center) and $y = |E_3 - E_4|/m_K$ (right)
of the selected NA48 $\klpigg$ candidates.
\label{fig:klpigg_m12_m34_y} }
\end{figure}

The invariant mass $m(\gamma_1 \gamma_2)$ of the two photons originating from the $\pi^0$ is shown
in Fig.~\ref{fig:klpigg_m12_m34_y} (left). NA48 finds 2558 $\klpigg$ candidates with an estimated background
of $82 \pm 12$ events, which are used for the branching ratio measurement.
For the fit of the VMD parameter $a_V$, 345 ambigous events which allow two possible $\pi^0 \to \gamma \gamma$
assignments are excluded.
The parameter $a_V$ is fitted using the distributions of both kinematic decay variables
$m(\gamma_3 \gamma_4)$ and $y = |E_3 - E_4|/m_K$ (Fig.~\ref{fig:klpigg_m12_m34_y}).
The result of the fit is
\[
a_V \: = \: -0.46 \pm 0.03_{\subrm{stat}} \pm 0.04_{\subrm{syst}} \hspace{1cm} {\rm (NA48, 2002)},
\]
with the systematic error being dominated by uncertainties of the acceptance evaluation ($\pm 0.03$)
and the parametrization of the $\klpipipi$ vertex\cite{bib:kambor} ($\pm 0.02$).

\begin{figure}[t]
\begin{center}
\epsfig{file=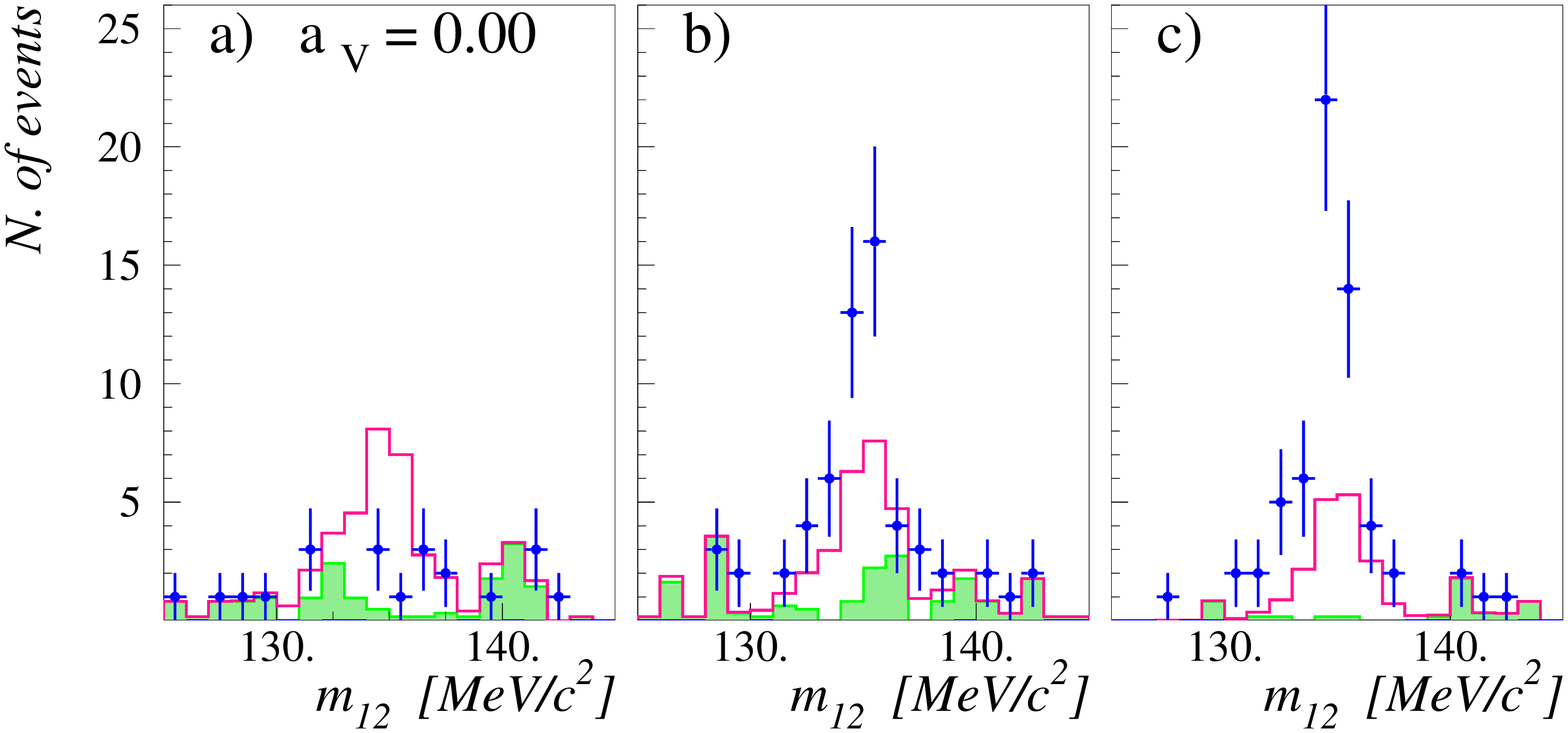,width=0.85\textwidth,height=0.23\textheight}
\epsfig{file=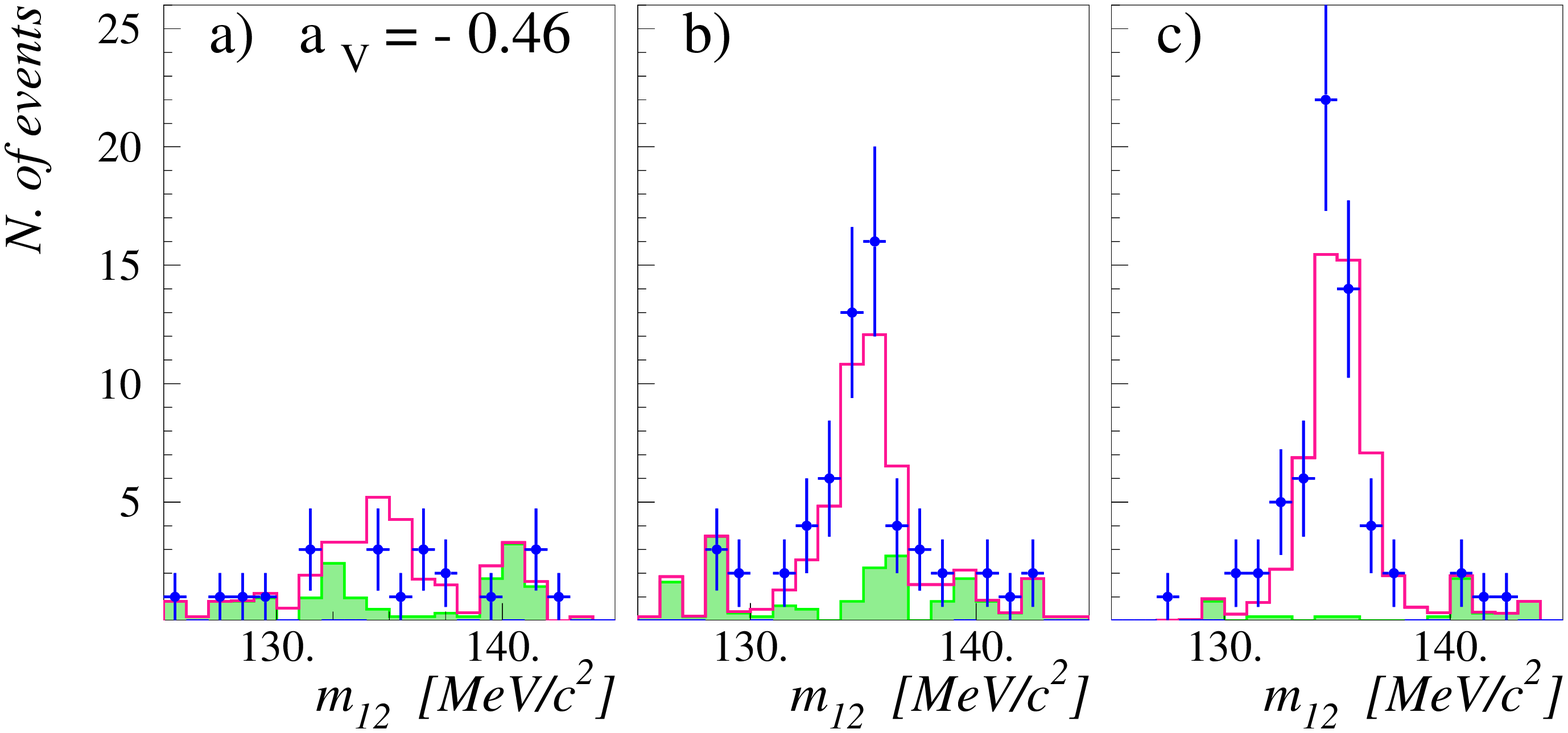,width=0.85\textwidth,height=0.23\textheight}
\end{center}
\vspace{-6mm}
\caption{\it NA48 data (crosses) and expected background (shaded) for 
a)  $30$~MeV/$c^2 <  m(\gamma_3 \gamma_4) < 110$~MeV/$c^2$, 
b) $160$~MeV/$c^2 <  m(\gamma_3 \gamma_4) < 240$~MeV/$c^2$, and 
c) $220$~MeV/$c^2 <  m(\gamma_3 \gamma_4) < 260$~MeV/$c^2$ 
together with the expectation of $a_V = 0.0$ (top) and $a_V = 0.46$ (bottom).
\label{fig:klpigg_av}}
\end{figure}

For low $m(\gamma_3 \gamma_4)$ the distributions of the $\pi^0$ signal $m(\gamma_1 \gamma_2)$
are shown in Fig.~\ref{fig:klpigg_av}. There is no significant signal of $\klpigg$ events for
$m(\gamma_1 \gamma_2) < 110$~MeV/$c^2$, as expected for $a_V \approx -0.46$ due to cancellation effects.
Nevertheless, in the region $160 < m(\gamma_3 \gamma_4) < 240$~MeV/$c^2$ a clear signal shows up,
which is evidence for a sizeable ${\cal O}(p^6)$ contribution to the $\klpigg$ amplitude.

\begin{figure}[t]
\begin{center}
\epsfig{file=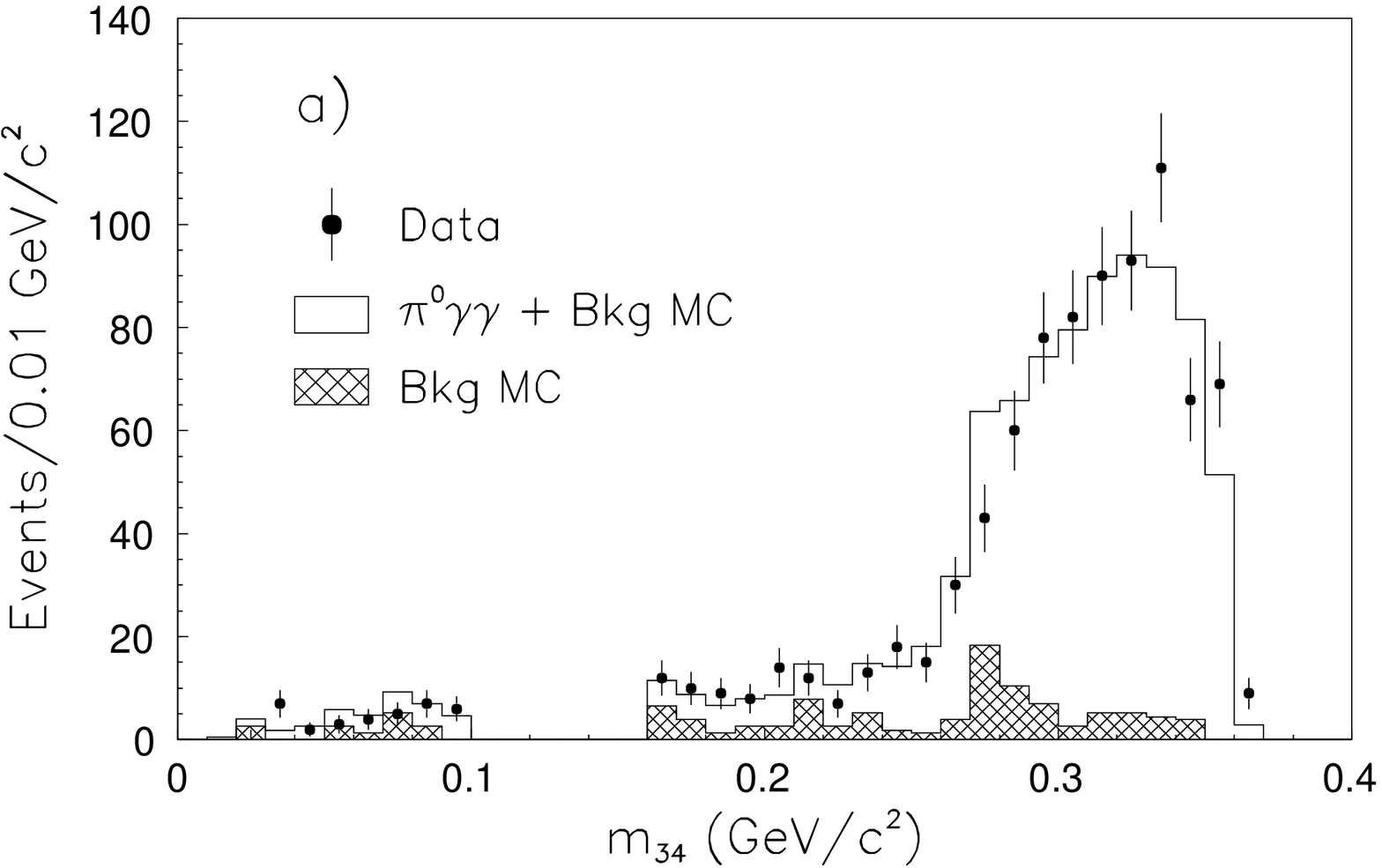,width=0.49\textwidth}
\epsfig{file=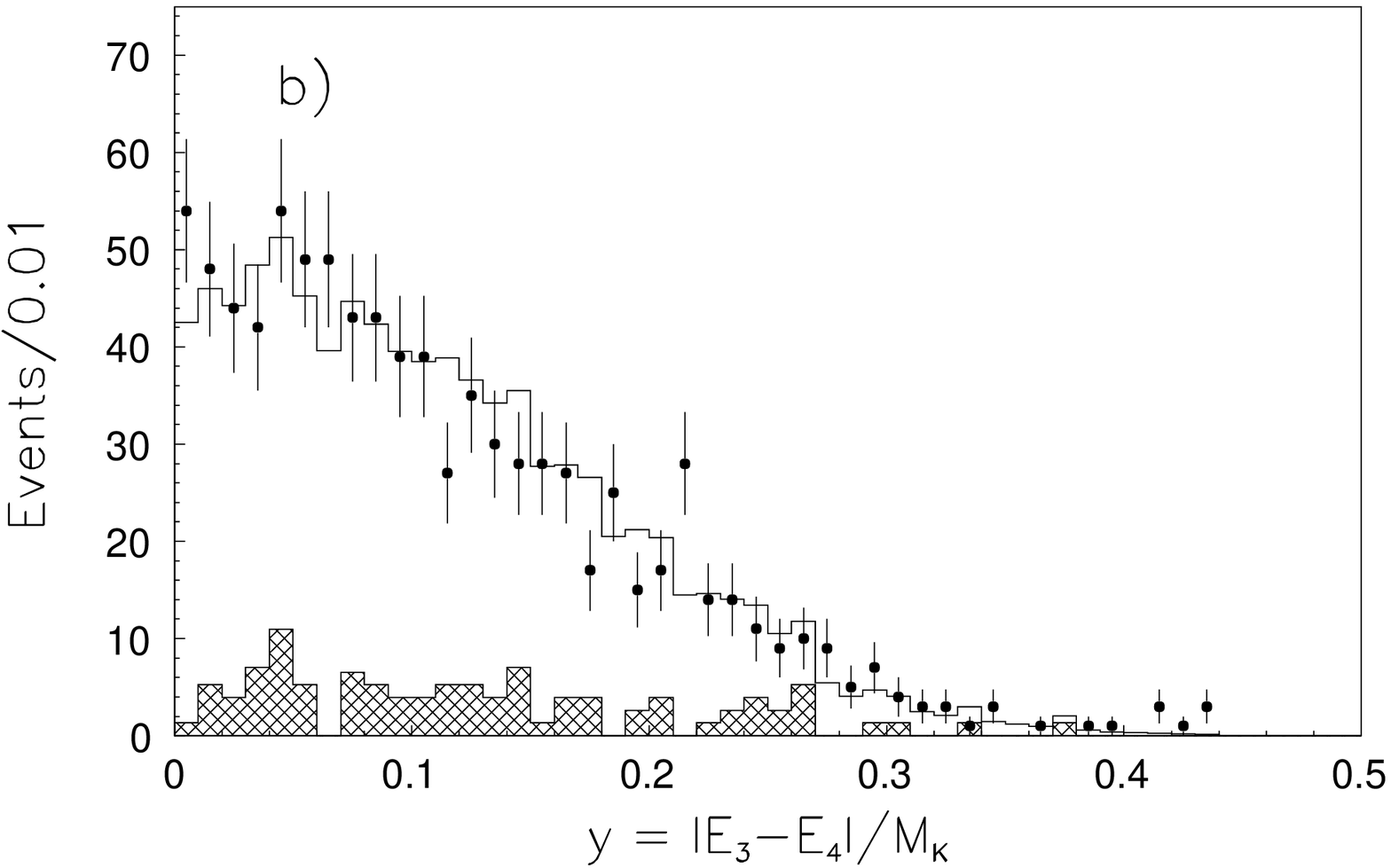,width=0.49\textwidth}
\end{center}
\vspace{-6mm}
\caption{\it Distributions of $m(\gamma_3 \gamma_4)$ (left) and $y = |E_3 - E_4|/m_K$ (right)
of the selected $\klpigg$ candidates of the KTeV experiment.
\label{fig:klpigg_ktev_m34_y} }
\end{figure}

However, an analysis of the KTeV experiment using the data of the years 1996 and 1997,
comes to a different result\cite{bib:klpigg_ktev}.
Performing a similar analysis as described above, KTeV finds 884 signal candidates with
estimated $111 \pm 12$ background events.
Fitting both the $m(\gamma_3 \gamma_4)$ and $y$ distributions (Fig.~\ref{fig:klpigg_ktev_m34_y})
the measured value of $a_V$ is
\[
a_V \: = \: -0.72 \pm 0.05_{\subrm{stat}} \pm 0.06_{\subrm{syst}} \hspace{1cm} {\rm (KTeV, 1999)},
\]
also measuring a large ${\cal O}(p^6)$ contribution,
but about $2.8$~standard deviations different from the NA48 result.
The systematics are dominated by the knowledge of the $3 \pi^0$ background.
While NA48 has finished the analysis on $\klpigg$, 
the KTeV collaboration plans to additionally analyze the data from the 1999 data taking period,
which might help to understand the different results of the two experiments.

The two experiments also arrive at different measurements of the $\klpigg$ branching fraction,
which can partially be explained by the strong dependence of the detector acceptances on $m(\gamma_3 \gamma_4)$:
\begin{eqnarray*}
{\rm NA48} \; (2002): \quad \Br(\klpigg) & = & ( 1.36 \, \pm \, 0.03_{\subrm{stat}} \, \pm \, 0.04_{\subrm{syst}} ) \times 10^{-6} \\*[2mm]
{\rm KTeV} \; (1999): \quad \Br(\klpigg) & = & ( 1.68 \, \pm \, 0.07_{\subrm{stat}} \, \pm \, 0.08_{\subrm{syst}} ) \times 10^{-6}
\end{eqnarray*}

The NA48 $a_V$ measurement indicates a negligible CP-conserving amplitude of the (yet to be observed)
direct CP-violating decay $\klpiee$\cite{bib:klpiee_theo}.
However, this does not hold for the $a_V$ value measured by KTeV.

\section{$\kspigg$}


The decay $\kspigg$ is dominated by the pion pole in $\kspizpiz$.
To be able to distinguish this decay from the dominating $\kspizpiz$, a minimum two-gamma invariant
mass of $z = m(\gamma_3 \gamma_4)^2/m_K^2 > 0.2$ is required.
One theoretical investigation based on ChPT exists\cite{bib:klpigg_theo}, predicting the shape
of the $m(\gamma_3 \gamma_4)$ distribution together with a branching fraction of
$\Br(\kspigg)|_{z \ge 0.2}  =  3.8 \times 10^{-8}$.
No experimental observation or limit has been published so far.

Using the data of a two-day high intensity $\ks$ run in 1999,
the NA48 experiment has performed a search for the decay $\kspigg$.
In this run about $3 \times10^8$ $\ks$ decays took place in the fiducial detector volume.

Because of the smallness of the expected branching fraction, the 
main experimental problem is the background suppression.
Any non-$\gamma$ activity in the photon anti-counters, drift chambers, and hadron calorimeter is vetoed.
To suppress $\kspizpiz$ events, the $\chi^2_{2\pi^0}$ of the event for being a $\kspizpiz$ decay is required to be larger
than 2000. In this way, only $0.1 \pm 0.1$ mismeasured $\kspizpiz$ events are expected, as determined from Monte Carlo simulation.
The main background then comes from $\kspizpiz$ decays with one lost and one
accidentally in-time photon. This background is estimated by simulated and randomly triggered events
to $2.1 \pm 0.1$ events.
The irreducible background coming from $\klpigg$ decays accounts for another $0.1$ expected background event.

\begin{figure}[t]
\begin{center}
\epsfig{file=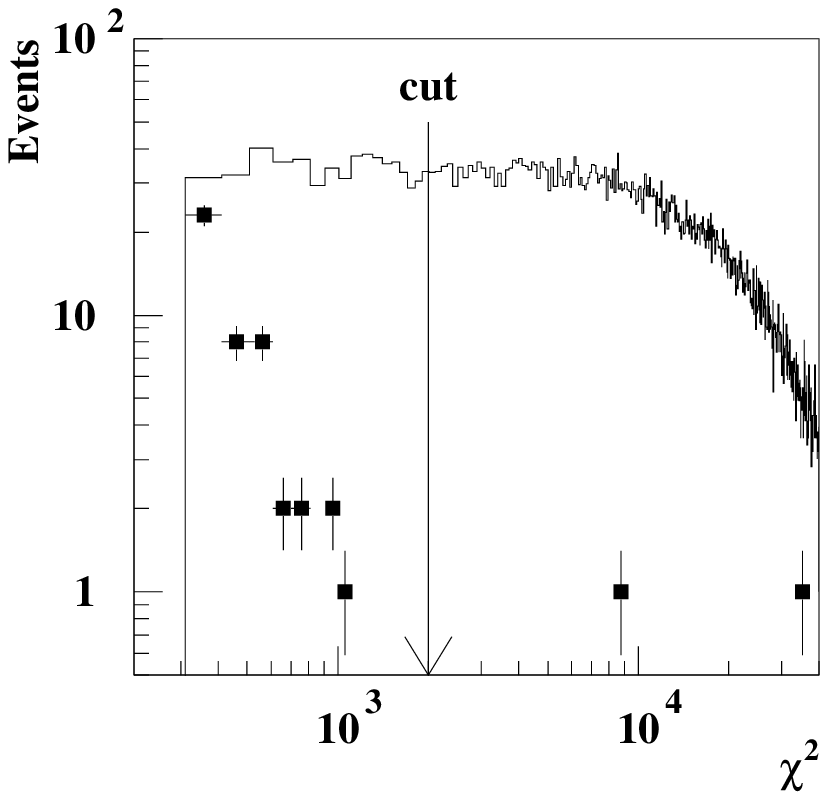,width=0.43\textwidth}
\end{center}
\vspace{-6mm}
\caption{\it Right: $\chi^2_{2 \pi^0}$ distribution of $\kspigg$ candidates under $\kspizpiz$ hypothesis (crosses).
The histogram shows the expected signal distribution.
\label{fig:kspigg_chi2}}
\end{figure}

The $\chi^2_{2\pi^0}$ distribution of the remaining events is shown in Fig.~\ref{fig:kspigg_chi2}.
Only two events above $\chi^2_{2\pi^0} = 2000$ survive the selection, compatible with the background
expectation of $2.3 \pm 0.2$ events.
Using $\kspizpiz$ decays for normalization, the limit on the branching fraction set by NA48 is
\[
\Br(\kspigg)|_{z \ge 0.2} \; < \; 4.4 \times 10^{-7} \quad {\rm at} \; 90\% \; {\rm CL.}
\]
This is the first limit set on the decay width of $\kspigg$.
However, it is still about one order of magnitude above the theoretical expectation.

\end{document}